\documentclass[twocolumn,preprintnumbers,amsmath,amssymb]{revtex4-1}
\usepackage{graphicx}
\usepackage{dcolumn}
\usepackage{bm}
\usepackage[usenames,dvipsnames]{xcolor}

\begin{document}
\preprint{v2.4}

\title{\color{red} Magnetoelastic instability in soft thin films}

\author{M.Poty}
\author{F.Weyer}
\author{G.Grosjean}
\author{G.Lumay}
\author{N.Vandewalle}
\address{GRASP, Institute of Physics B5a, University of Li\`ege, B4000 Li\`ege, Belgium.}

\begin{abstract}
Ferromagnetic particles are incorporated in a thin soft elastic matrix. A lamella, made of this smart material, is studied experimentally and modeled. We show herein that thin films can be actuated using an external magnetic field applied through the system. The system is found to be switchable since subcritical pitchfork bifurcation is discovered in the beam shape when the magnetic field orientation is modified. Strong magnetoelastic effects can be obtained depending on both field strength and orientation. Our results provide versatile ways to contribute to many applications from the microfabrication of actuators to soft robotics. As an example, we created a small synthetic octopus piloted by an external magnetic field.
\end{abstract}

\maketitle


\section{Introduction}

The elastic moduli of soft materials can be drastically modified by the presence of inclusions such as solid particles \cite{particles}, nanotubes \cite{nanotubes} or even holes \cite{holes}. It has been also proposed to consider magnetic inclusions in order to change the bulk properties of the materials \cite{bulk} by the application of an external field. Such smart materials sensitive to external stimuli and fields represent opportunities for microfabrication.  Recently, thin elastic films  in which aligned magnetic particles are included, were proposed as swimming robots \cite{sitti_apl}, soft actuators or valves \cite{velev}.

In the present paper, our main motivation is to create thin soft elastic films sensitive to the presence of a magnetic field in order to study and rationalize magnetoelastic effects. In the section \ref{sec:methods}, the fabrication method will be detailed. Then, thin films will be experimentally studied and data collected. The experimental results evidence a magnetoelastic instability that will be described in section \ref{sec:model}. Finally, we provide some application of magnetoelastic lamella at the centimeter scale.

\section{Methods}
\label{sec:methods}

\begin{figure}[h]
\begin{center}
\vskip 0.2 cm
\includegraphics[width=8.5cm]{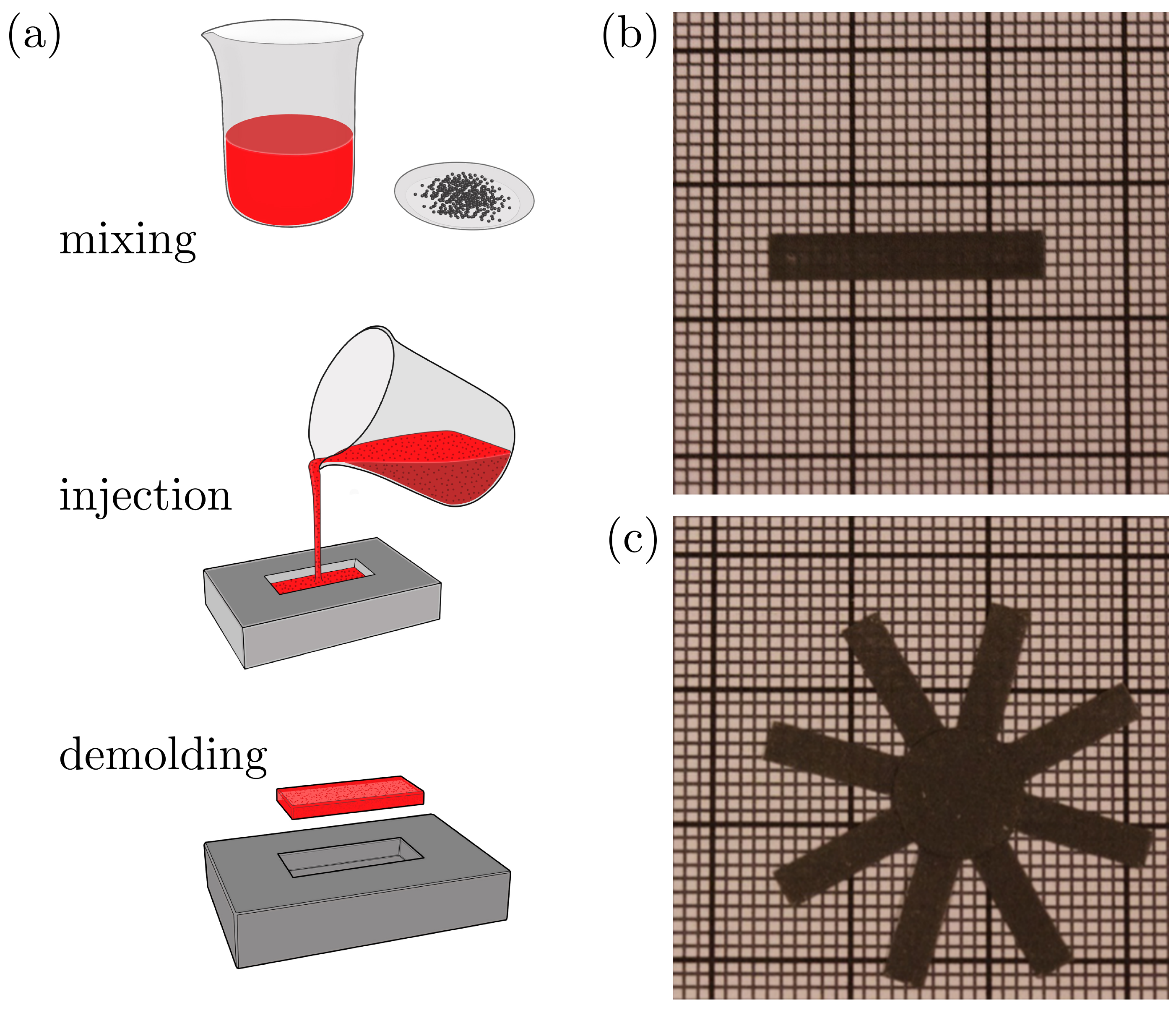}
\vskip -0.2 cm
\caption{(a) Illustration of the fabrication process of a soft magnetoelastic beam. (b) Picture of a thin elastic sheet having a rectangular shape. The scale of this lamella is given by the underlying millimeter paper. (c) Picture of a thin elastic sheet with a star shape. The scale is given by the underlying millimeter paper. }
\label{sketch}
\end{center}
\end{figure}

The fabrication process is illustrated in Figure \ref{sketch}(a). We incorporate ferromagnetic particles in a liquid vinylpolysiloxane phase before reticulation. Those particles are iron oxide aggregates with a typical diameter ranging from 50 to 300 $\rm \mu m$. It has been proved \cite{geoffroy} that, in the presence of a magnetic field, the rheology of such powdered materials is drastically modified due to the dipole-dipole interaction between neighboring grains. The residual magnetization of this powder is however weak. The polymer/particule mixture is then inserted in a thin mold for obtaining rectangular magnetoelastic sheets after polymer reticulation, as shown in Figure \ref{sketch}(b). Typical thickness is 350 $\rm \mu m$. Thin objects with complex shapes can also be obtained, like the star-like structure with 8 branches shown in Figure \ref{sketch}(c). This system will be considered at the end of this paper.
 
The dispersion of the particles is homogeneous in the elastic sheet. Various densities $\eta$ of particles were considered for the elastic matrix, characterized by a typical Young modulus around $Y=0.3 \, {\rm MPa}$ \cite{bob}. Since high fractions lead to strong effects, we present in the following the results for $\eta = 0.75 \, {\rm mg/cm^3}$ only.

\section{Experimental results}
\label{sec:experimental}

\begin{figure}[h]
\begin{center}
\vskip 0.2 cm
\includegraphics[width=8.5cm]{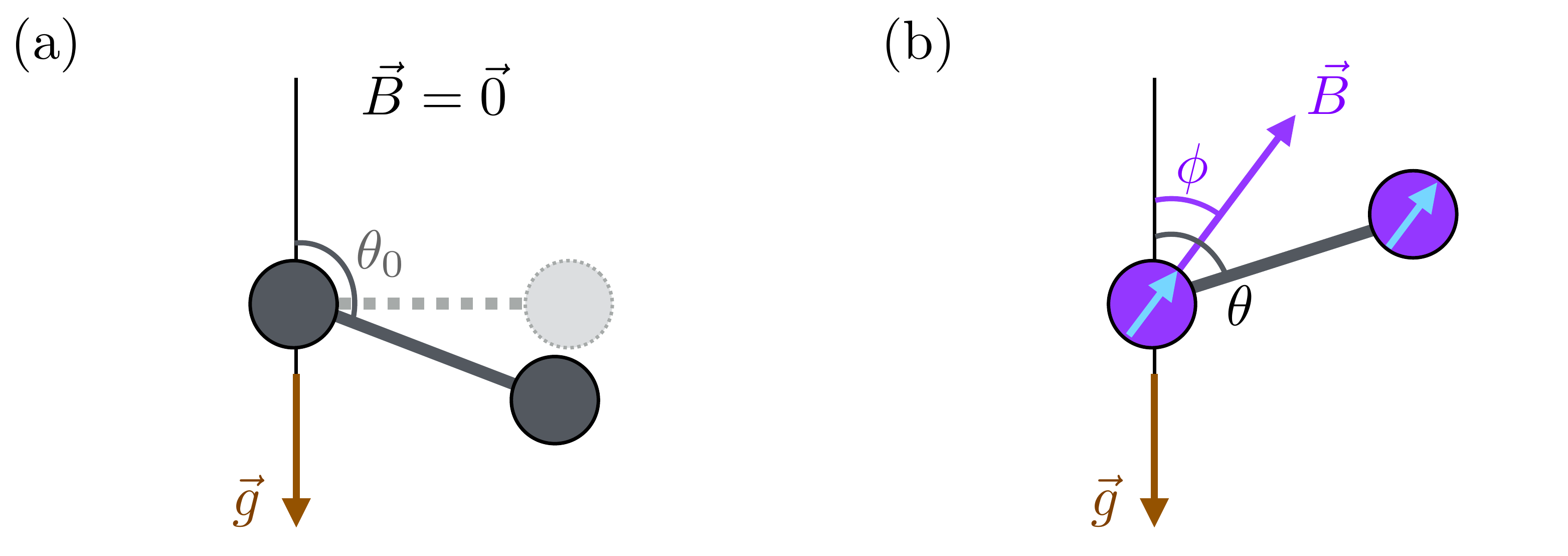}
\vskip 0.2 cm
\includegraphics[width=8.5cm]{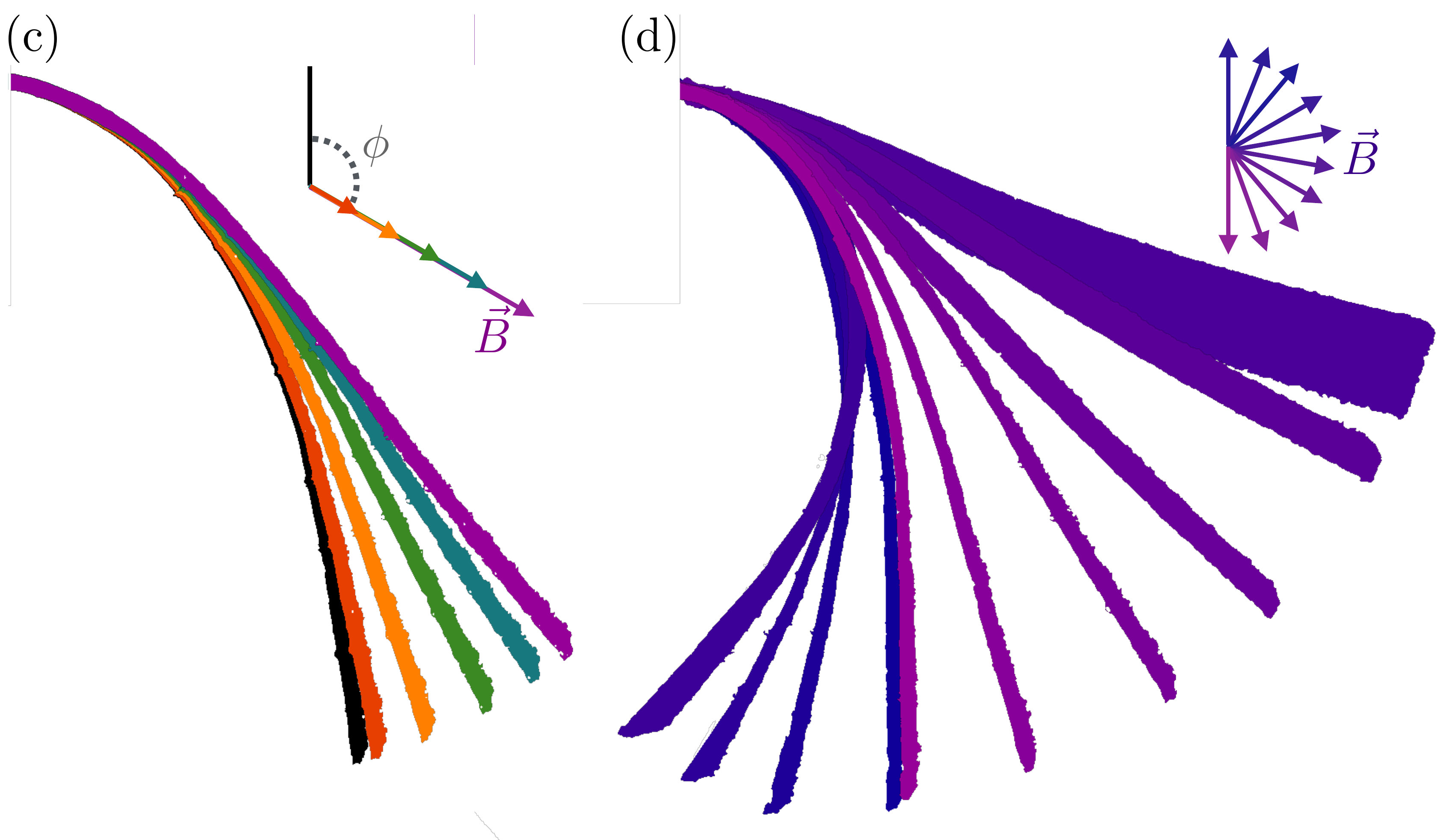}
\vskip -0.2 cm
\caption{(a) Two particles linked by an elastic rod fixed at the left. Since gravity acts on the free extremity (second particle), the equilibrium angle $\theta_0$ is larger than $90^\circ$. (b) Sketch of the same system (a) when a magnetic field $\vec B$ making an angle $\phi$ with the vertical axis is applied to the system. Induced dipoles modify the orientation of the elastic rod from $\theta_0$ to $\theta$. (c) Superimposed images of the beam as a function of the magnetic field strength from 0 G to 105 G by steps of 21 G . Different colors illustrate the field strength. The field makes an angle $\phi = 120^\circ$ with the vertical axis. The length of the magnetoelastic film is 15 mm. (d) The same cantilever beam for a field of 105 G when the angle $\phi$ rotates from $0^\circ$ to $180^\circ$.  }
\label{snapshots}
\end{center}
\end{figure}

By using an external field $\vec B$, the particles of the flexible film become induced dipoles. As a result, the dipole-dipole interactions may affect the shape of the elastic sheet due to some competition between elastic and magnetic energies. This phenomenon can be intuitively captured by the sketches of Figure \ref{snapshots}(a) and \ref{snapshots}(b). The former sketch shows two particles linked by an elastic rod of length $r$. The left particle is fixed while the second one is attached at the free extremity of a horizontal rod. Gravity acts against elastic energy such that an equilibrium angle $\theta_0$, different from $90^\circ$, is observed. When the field is switched on, as shown in Figure \ref{snapshots}(b), particles act as dipoles $\vec \mu$ oriented along the field making an angle $\phi$ with the vertical axis. The dipoles tend to form chains along the field orientation. These chains possess their own elastic behavior due to the magnetism \cite{vella,messinaSM,messinaEPJE,vdw}. This orientation effect is understood by considering the magnetic potential energy of both dipoles \begin{equation}
U= {\mu_0 \mu^2 \over 4\pi r^3} \left[ 1-3 \cos^2 (\theta-\phi) \right]
\label{eq:dipoles}
\end{equation}
which is minimized when $\theta-\phi=0$, i.e. when the rod is oriented along the magnetic field.  As a result, the new equilibrium angle $\theta$ when $B\ne 0$ can be completely different from $\theta_0$. 

In order to study the various effects of an external field on such compound films, we built a system of Helmholtz coils, producing a uniform magnetic field, which can rotate around a beam cantilever made of this soft material. The materials have been studied in air or immersed in water. The later situation was chosen to reduce gravity effect acting on the beam, thanks to buoyancy. 

Figure \ref{snapshots} shows two different sets of experiments for the same lamella. The first picture, i.e. Figure \ref{snapshots}(c), is a superposition of snapshots obtained when the field increases from 0 G to 105 G by steps of 21 G. The field $\vec B$ makes an angle of $\phi=120^\circ$ with the vertical axis, as indicated by the arrows. Different colors, from red to violet, correspond to different field strengths. One observes a significant effect : the beam orientation changes with field strength $B$. The free extremity of the beam cantilever tends to adopt the field angle $\phi$, as expected from Eq.(\ref{eq:dipoles}). For this beam, the dipole-dipole interaction between particles is strong enough to compete with gravity and elastic energies. Since $\eta \propto r^{-3}$, the magnetoelastic effect, as reported here, depends linearly on the concentration of particles in the beam cantilever. 

The second picture, shown in Figure \ref{snapshots}(d), is also a superposition of snapshots for the same system when the field strength is kept constant at the maximum value $B=105$ G while the orientation of the field is modified clockwise from $\phi=0^\circ$ to $\phi=180^\circ$ by steps of 20$^\circ$. A strong effect is observed when the field orientation reaches a particular angle, corresponding nearly to the field perpendicular to the beam. There, the angle $\theta$ between the beam end and the vertical axis jumps by about 60$^\circ$ around this critical point. The effect comes from the anisotropic nature of the dipolar magnetic interaction, as we will see below. 

From the above observations, we can conclude that it is possible to induce a magnetoelastic effect in materials with high enough concentration of ferromagnetic inclusions. This effect depends on both field strength and orientation. Let us now describe more quantitatively those effects.

\begin{figure}[h]
\begin{center}
\vskip 0.2 cm
\includegraphics[width=8cm]{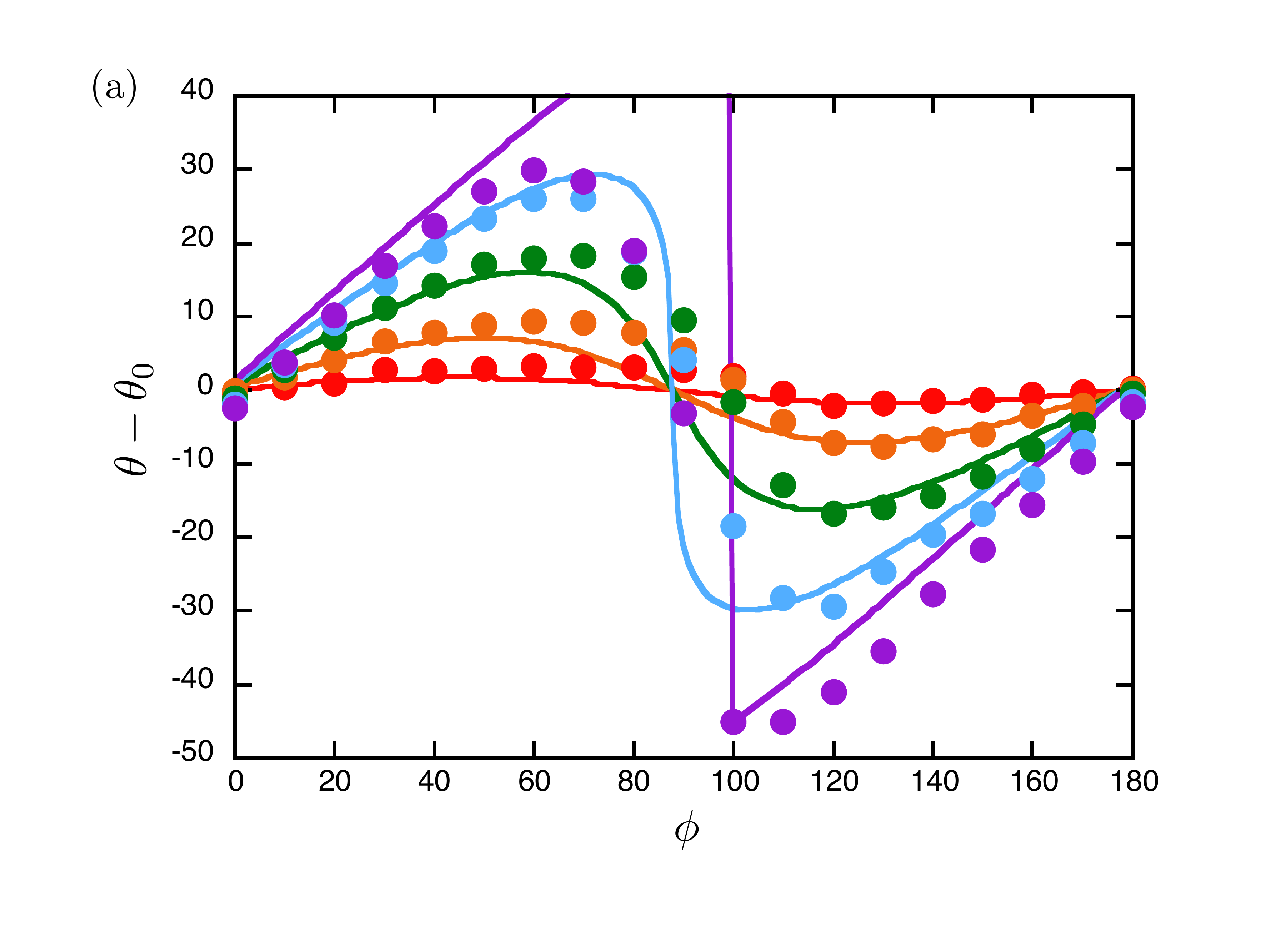}
\includegraphics[width=8cm]{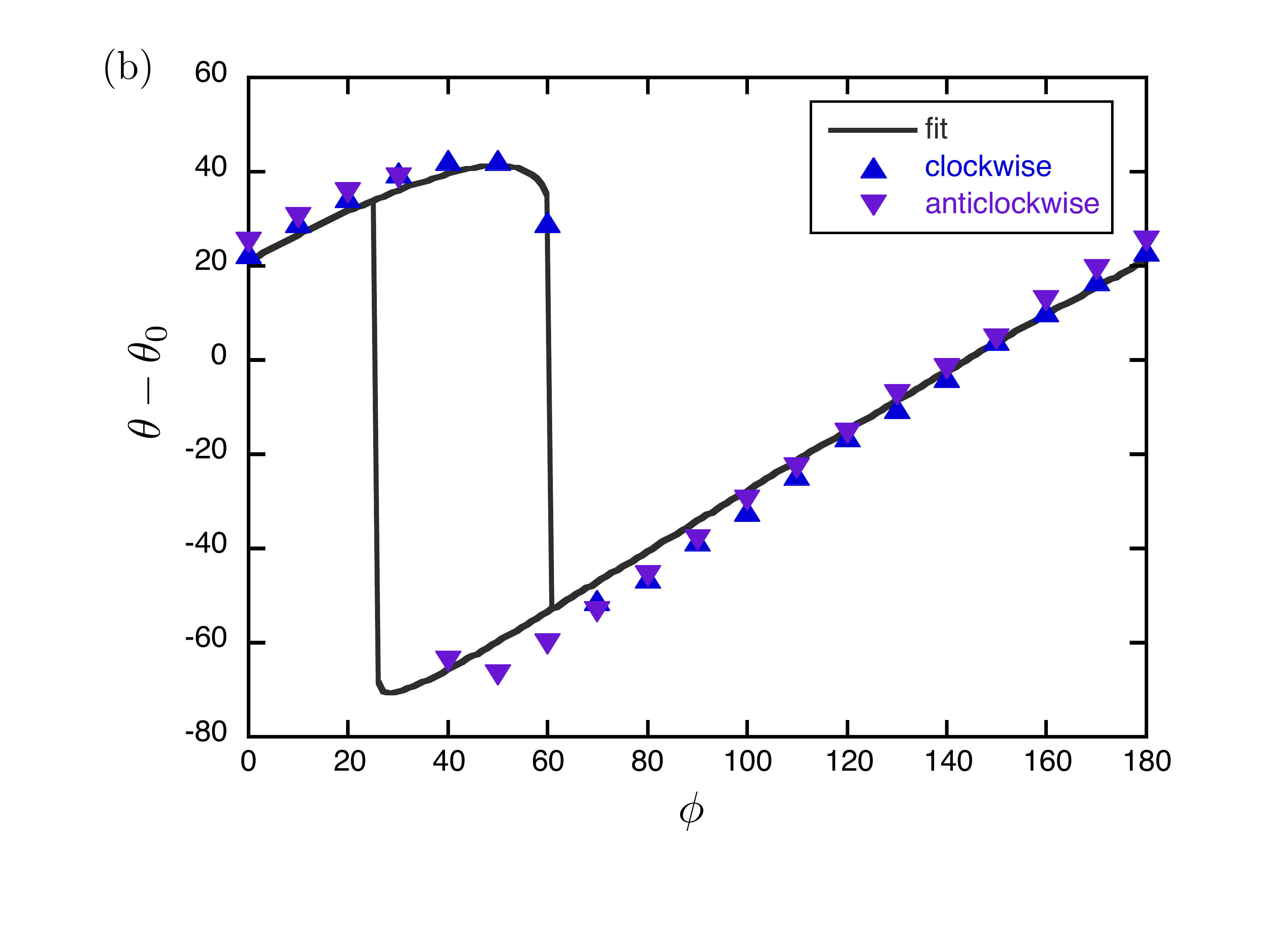}
\vskip -0.2 cm
\caption{(a) Dots present the measurements of the angle deviation $\theta-\theta_0$ of the beam end as a function of field orientation $\phi$. The different colors corresponds to different field strengths from 21 G to 105 G by steps of 21 G (same color code as Figure \ref{snapshots}(c)). Curves are fits using the model based on Eq.(2). (b) For a similar beam, a strong field rotates clockwise and above a critical angle of about $\phi=60^\circ$ the system jumps abruptly to another configuration. When the field then rotates anti-clockwise, the jump takes place at a lower critical angle, i.e. around $40^\circ$. The angle difference is a signature of hysteresis. Curves are fits using the model explained in the main text. }
\label{results}
\end{center}
\end{figure}

By image analysis, we have determined the angle $\theta$ at the end of the beam on all the pictures. Without a magnetic field, the angle is $\theta_0$. The data are reported in Figure \ref{results}(a). Each curve is an experiment where the field orientation changes. Five different values of the field strength are illustrated, using the same color code as Figure \ref{snapshots}(c).   The angle deviation $\theta-\theta_0$ of the beam is reported as a function of the field orientation $\phi$.  By increasing the strength, one observes that the angle oscillates around the equilibrium value $\theta_0$. For high field strengths, the oscillation becomes a discontinuous jump. This discontinuity is the signature of a subcritical pitchfork bifurcation, that we will evidence below. The critical angle seems to depend on the relative position between the cantilever and the external field. 

Figure \ref{results}(b) presents a particular set of data for another magnetoelastic film. In this second experiment, the field has been rotated clockwise and then counterclockwise. For a similar beam, when a strong field rotates clockwise, the system jumps to another configuration above a critical angle of about $60^\circ$ t. This is a discontinuous jump. When the field then rotates anti-clockwise, the abrupt jump takes place at a lower critical angle below $40^\circ$. Therefore, a hysteresis loop takes place in the system.

\section{Model}
\label{sec:model}

In order to model the magnetoelastic effect reported hereabove, we propose the minimal system with a low set of variables exhibiting the level of complexity found in our experiments \cite{anderson}. Let us consider the elastic rod connecting two dipoles presented in Figure \ref{snapshots}(a). When both gravity and external magnetic field are considered, the dimensionless energy of the system can be given by
\begin{equation}
u= \left( \theta-90 \right)^2 + \mathcal{G} \cos \theta + \mathcal{M} \left[1-3\cos^2 \left(\phi - \theta  \right) \right]
\label{eq:u}
\end{equation}
where the numbers $\mathcal{G}$ and $\mathcal{M}$ are respectively the gravitoelastic and magnetoelastic numbers.

The gravitoelastic number $\mathcal{G}$ measures the competition between gravity and elastic energies. It is expected to be 
\begin{equation} 
\mathcal{G} = {\rho g \ell \over Y}, 
\end{equation}
where $\ell$ is a characteristic length/tickness of the system. In order to bend the elastic rod, gravity effects should be of the same order of magnitude as the first term of Eq.(2), i.e. elastic deformations appear above $\mathcal{G} \approx 10$, as obtained below. 

The magnetoelastic number $\mathcal{M}$ measures the competition of magnetic and elastic energies. The dipoles $\mu= {\chi B a^3 \over \mu_0}$ are expected to appear on the particles of volume $a^3$. The iron oxide material is ferromagnetic with a susceptibility $\chi$ above 100 \cite{handbook}. Moreover, our particules are far from spherical. Due to demagnetization effects, the various shapes of the particles affect deeply the effective susceptibility, and we assume that the effective susceptibility is around 100. At high concentration, the mean distance between neighboring particles is close to $a$, one has therefore 
\begin{equation} 
\mathcal{M} = {\chi^2 B^2 \over \mu_0 Y}.
\end{equation} 
For the materials and parameters of our experiment, we obtain roughly $M \approx 10$ at high fields, meaning that the magnetic energies have the same order of magnitude than elastic energies in our lamella.

\begin{figure}[h]
\begin{center}
\vskip 0.2 cm
\includegraphics[width=8.5cm]{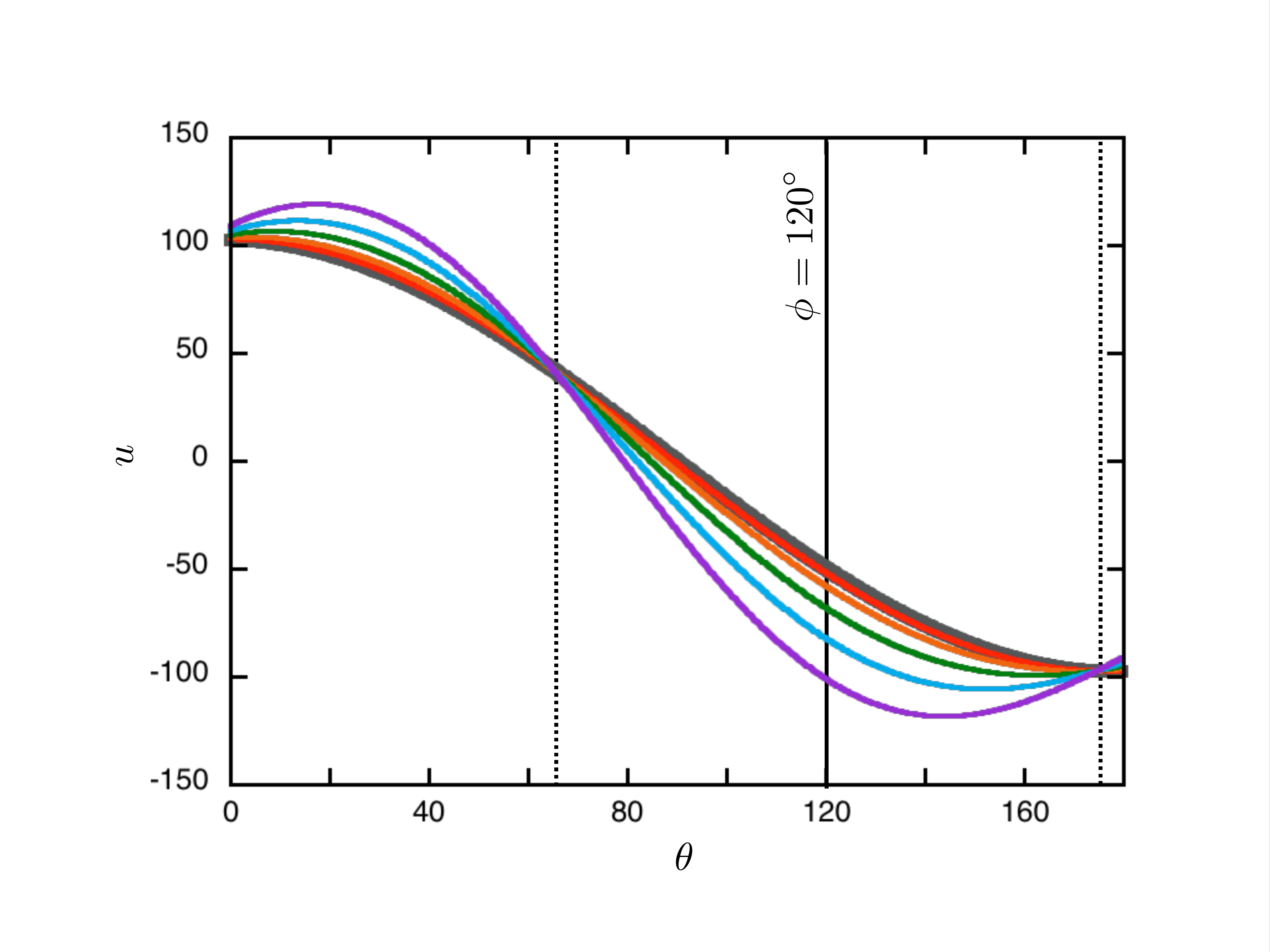}
\vskip -0.2 cm
\caption{ The dimensionless potential $u$ as a function of $\theta$. The dark curve corresponds to $B=0$. A field is applied with an angle $\phi=120^\circ$, denoted by the vertical line. The color code corresponds to increasing strength values of Figures \ref{snapshots}(c) and \ref{results}(a). The dashed vertical lines indicate the $\theta'$ values for which the field has no effect.   }
\label{model_theta}
\end{center}
\end{figure}

When $B=0$, i.e. when $\mathcal{M}=0$, a single minimum is found for $u(\theta)$ determining the equilibrium angle $\theta_0$. This equilibrium angle $\theta_0 \ge \pi/2$ under gravity depends solely on $\mathcal{G}$. The thick dark curve in Figure \ref{model_theta} presents the potential $u(\theta)$ as obtained for flexible films for $\mathcal{G}=100$ and $\mathcal{M}=0$. The minimum $\theta_0$ is close to $180^\circ$ in order to mimic the lamella of Figure \ref{snapshots}(c). When the external field increases, $\mathcal{M} \propto B^2$ becomes nonzero, the minimum of the potential $u$ is seen to shift towards $\phi$. In Figure \ref{model_theta}, an increasing field is applied with an angle $\phi=120^\circ$. The case $\phi=\theta=120^\circ$ is denoted by a vertical line in this Figure. Following the evolution of $u$ when $B$ increases, one realizes indeed that the lamella tends to be oriented along $\vec B$. Dashed vertical lines indicates specific angles $\theta'$ for which $1-3\cos(\phi-\theta')^2=0$ where the field seems has no effect : all curves cross at these specific angles.

In the last term of Eq.(\ref{eq:u}), one notices that the additional minima take place at $\phi-\theta= 0^\circ$. By changing $\phi$, the equilibrium position $\theta$ is modified and oscillates around $\theta_0$. The data of Figure \ref{results}(a) are fitted by the model with a single dimensionless $\mathcal{G}=100$ value for all curves, and a constant $\mathcal{M}/B^2=0.0023 \, {\rm G^{-2}}$. The agreement between the model and the experimental data is remarkably good. However, deviations are seen in particular for high field strengths. It should be recalled that the model only considers two dipoles. Of course, a more elaborated model can be proposed to capture the details of the film behavior. However, it should be based on the multiple interactions between many dipoles leading to complex theoretical developments as found recently for chains of magnets \cite{vella,vdw,messinaSM,messinaEPJE}. 

\begin{figure}[h]
\begin{center}
\vskip 0.2 cm
\includegraphics[width=8.5cm]{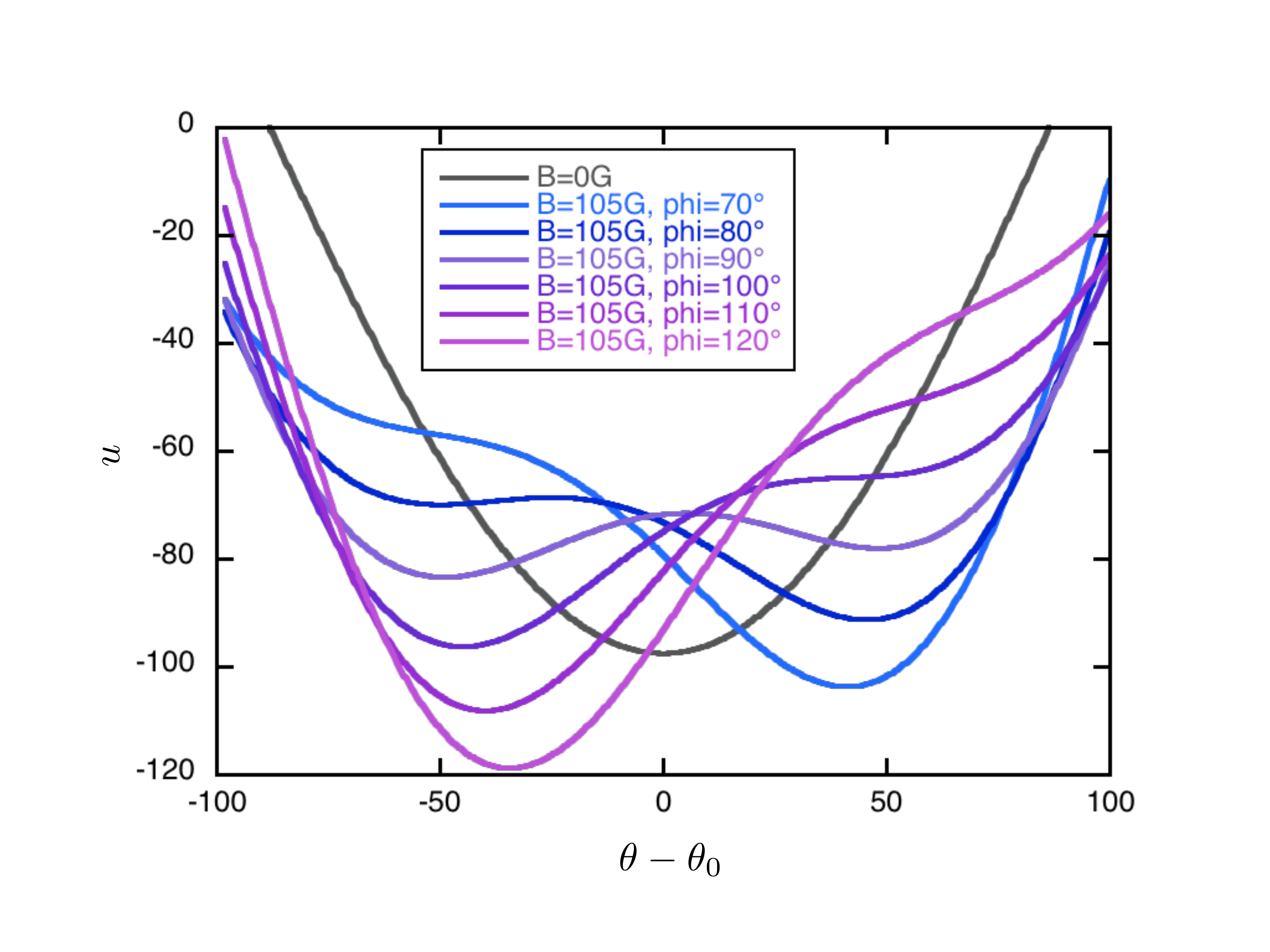}
\vskip -0.2 cm
\caption{The dimensionless potential $u$ as a function of $\theta$ around the zero field equilibrium value $\theta_0$, as given by Eq.(\ref{eq:u}). The dark curve corresponds to the case $B=0$. Colored curves show how the potential is modified when a strong field is applied along various orientations. The color code corresponds to Figure \ref{snapshots}(d). When the field angle $\phi$ increases, the local minimum becomes more and more separated from $\theta_0$. At some point ($\phi=100^\circ$), the equilibrium situation jumps to another minimum. A discontinuity is therefore expected, as reported in our Figures 3(a) and 3(b).   }
\label{model_phi}
\end{center}
\end{figure}

When the field strength becomes more important, the system may jump from one minimum to a neighboring one, explaining the abrupt modification of the equilibrium angle and the jump at some $\phi$ value, as seen in Figure \ref{results}(b). This is a subcritical bifurcation, involving a memory effect and hysteresis. Let us explore the model to evidence such a jump. 
In Figure \ref{model_phi}, the potential $u$ is plotted around the zero field  equilibrium angle $\theta_0$. When the angle $\phi$ is increased, the global minimum is seen to increase accordingly till $\phi \approx 100^\circ$, where another minimum at lower $\theta$ values is observed in $u$. Therefore,  the system jumps abruptly between positive and negative $\theta-\theta_0$ values. Since two local minima are seen for a range of $\phi$ values, metastability is expected. As a consequence hysteresis loops are found.

Figure \ref{results}(b) shows a hysteresis loop obtained with this model fitting the data. The parameter $\mathcal{G}=3.15$ has first been fixed in order to show an angle $\theta_0$ close to the one of Figure \ref{results}(b). The low $\mathcal{G}$ value comes from a different equilibrium angle $\theta_0$ than experiments of Figure \ref{results}(a). Then, the parameter $\mathcal{M}=1.20$ has been fixed in order to show a hysteresis similar to the one reported in Figure \ref{results}(b). The model nicely reproduces the experimental data and the hysteresis loop. 

\section{Discussion and applications}
\label{sec:application}

Recent works considered also the inclusion of magnetic particles in elastic films \cite{velev,singh,sitti_apl}. Our work provides a versatile way to create complex objects and our experimental investigations prove the existence of an instability. The magnetoelastic soft material is seen to be switchable. Moreover, these behaviors seem to be captured by our two dipole model. The latter can be generalized in a future work  to series of dipoles in a long chain constituted of elastic / magnetic parts. 

\begin{figure}[h]
\begin{center}
\vskip 0.3 cm
(a) \includegraphics[width=8.0cm]{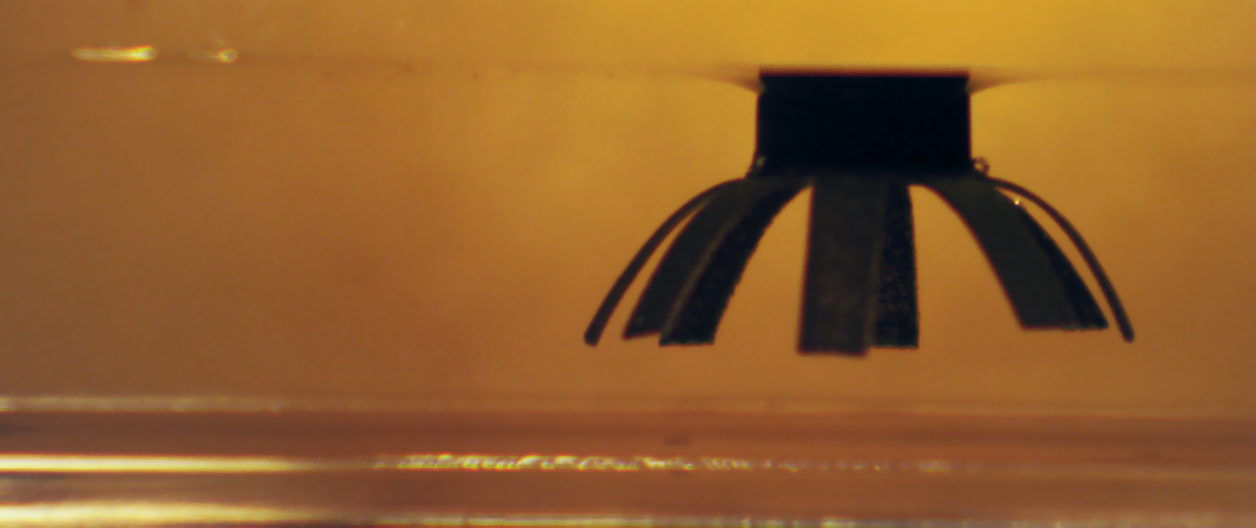}
(b) \includegraphics[width=8.0cm]{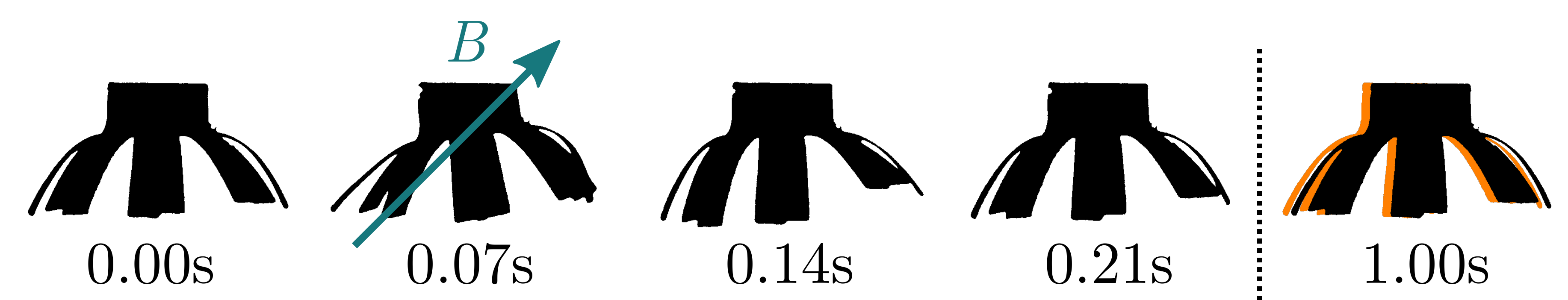}
\vskip -0.2 cm
\caption{(a) Picture of a small magnetic octopus floating below a liquid-air interface. The diameter of the cylinder is 1 cm. (b) Snapshots of the successive configurations of the octopus when a short pulse of an inclined magnetic field is applied on the second picture. The tentacles are moving according to our model. As a result, the entire system moved in the field direction, as illustrated in the last picture superimposing the initial position.  }
\label{spatio}
\end{center}
\end{figure}

Since the fabrication method is simple, it should be remarked that our magnetoelastic material can be used to create small object of any shape. By small, we mean that the gravity effect should be dominated by magnetoelastic effects, i.e. when $\mathcal{M}$ is non negligeable before $\mathcal{G}$. As an example of a possible application, we created an octopus by gluing the star shape magnetoelastic film of Figure \ref{sketch}(c) to a disk suspended at the liquid-air interface. This centimeter scale system is shown in Figure \ref{spatio}(a). An oscillating vertical field will provoke a synchroneous oscillation of the 8 tentacles. In this case, the mixing of fluids will be favored by oscillating tentacles driven by the field. Another motion, shown in Figure \ref{spatio}(b), can be provoked by a magnetic pulse : during a short period, a magnetic field inclined at $45^\circ$ is applied. The tentacles being nearly perpendicular to the field were strongly displaced, while the amplitude of the phenomenon is limited for the other tentacles. As a result, a translational motion of the entire system is provoked along the water-air interface, as emphasized in the last picture of Figure \ref{spatio}(b). Successive pulses allow for a magnetically driven motion. The control of the octopus motion is achieved by playing with the orientation of the intermittent field.

Since the fabrication method is simple, many applications at both mesoscopic and macroscopic scales can be found in various domains such as fluid transport, valves, motors using magnetic flagella, fluid mixing using synthetic cilia \cite{cilia} or other soft robotics issues \cite{softrobotics} including reconfigurable robotics \cite{sitti}. The instability evidenced in our work may induce sharp effects of interest for future applications. Moreover, magnetic remanence could be also used to induce a memory of the initial shape as done in \cite{sitti_apl}. 

\section{Summary}

In summary, we created this elastic films in which ferromagnetic particles are dispersed. We studied their behavior when submitted to an external magnetic field. We evidenced a magnetoelastic  instability that we modeled. It can be used for actuating soft systems, as we demonstrated in a remote controlled motion experiment, therefore opening new possible applications for this class of soft materials.

\vskip 0.2 cm
{\bf Acknowledgments} -- This work is financially supported by the University of Li\`ege (grant FSRC-11/36). FW thanks the FNRS and GG thanks FRIA for financial support.



\begin{thebibliography}{99}

\bibitem{particles} T.T.Wu,  Int. J. Solids Structures {\bf 2}, 1-8 (1966)

\bibitem{nanotubes} D.Qian, E.C.Dickey, R.Andrews and T.Rantell, Appl. Phys. Lett. {\bf 76}, 2868 (2000)

\bibitem{holes} B.Florijn, C.Coulais, and M.van Hecke, Phys. Rev. Lett. {\bf 113}, 175503 (2014)

\bibitem{bulk} Z.Varga, G.Filipcsei, and M.Zrinyi, Polymer {\bf 47} 227-233 (2006)

\bibitem{sitti_apl} E.Diller, J.Zhuang, G.Z.Lum, M.R.Edwards and M.Sitti, Appl. Phys. Lett. {\bf 104}, 174101(2014)

\bibitem{velev} R.Mishra, M.D.Dickey, O.Velev and J.B.Tracy, Nanoscale {\bf 8}, 1309 (2016)

\bibitem{geoffroy} G.Lumay and N.Vandewalle, Phys. Rev. E {\bf 78}, 061302 (2008)

\bibitem{bob} N.Adami and H.Caps, EPL  {\bf 106}, 46001 (2014)

\bibitem{vella} D.Vella, E.du Pontavice, C.L.Hall and A.Goriely, Proc. R. Soc. Lond. A {\bf 470}, 2162 (2014)

\bibitem{messinaSM} I.Stankovic, M.Dasica and R.Messina, Soft Matter {\bf 12}, 3056-3065 (2016)

\bibitem{messinaEPJE} R.Messina, and L.Spiteri, Eur. Phys. J. E {\bf 39}, 81 (2016)

\bibitem{vdw} N.Vandewalle and S.Dorbolo, New J. Phys. {\bf 16}, 013050 (2014)

\bibitem{anderson} P.W.Anderson, Physics Today {\bf 41}, 9 (1988)

\bibitem{handbook} W.M.Haynes, Handbook of Chemistry and Physics, 93th Ed. (CRC Press, 2012) 

\bibitem{singh} K.Singh, C.R.Tipton, E.Han and T.Mullin, Proc. Roy. Soc. A {\bf 469}, 20130111 (2013)

\bibitem{cilia} J.den Toonder, F.Bos, D.Broer, L.Filippini, M.Gillies, J.de Goede, T.Mol, M.Reijme, W.Talen, H.Wilderbeek, V.Khatavkarb and  P.Anderson, Lab Chip {\bf 8}, 533 (2008)

\bibitem{softrobotics} R.F.Shepherd, F.Ilievski, W.Choi, S.A.Morin, A.A.Stokes, A.D.Mazzeo, X.Chen, M.Wang and G.M.Whitesides, PNAS {\bf 108}, 20400 (2011)

\bibitem{sitti} E.Diller, N.Zhang, M.Sitti, J. Micro-Bio Robot {\bf 8}, 121 (2013)

\end{thebibliography}
\end{document}